\documentclass[letterpaper, 10 pt, conference]{ieeeconf}  

\IEEEoverridecommandlockouts           

\usepackage{cite}
\usepackage{amsmath,amssymb,amsfonts}

\usepackage{algorithmic}
\usepackage{graphicx}
\usepackage{textcomp}
\usepackage{xcolor}
\def\BibTeX{{\rm B\kern-.05em{\sc i\kern-.025em b}\kern-.08em
    T\kern-.1667em\lower.7ex\hbox{E}\kern-.125emX}}
\usepackage[utf8]{inputenc}
\usepackage{amsmath,graphicx}
\usepackage{mathrsfs}
\usepackage{xcolor}
\usepackage{mathtools}
\usepackage{enumerate}
\usepackage{booktabs}
\usepackage{romannum}
\usepackage{float}
\usepackage{amsthm}
\newtheorem{theorem}{Theorem}
\newtheorem{lm}{Lemma}

\makeatletter
\renewcommand*\env@matrix[1][\arraystretch]{%
  \edef\arraystretch{#1}%
  \hskip -\arraycolsep
  \let\@ifnextchar\new@ifnextchar
  \array{*\c@MaxMatrixCols c}}
\makeatother

\newcommand{\argminD}{\arg\!\min} 
\newcommand{\argmaxD}{\arg\!\max} 

\newcommand{\finite}{\star}

\newcommand{\RR}{{\mathbb{R}}}

\DeclareUnicodeCharacter{2212}{-}
\title{\LARGE \bf
Guarding a Translating Line with an Attached Defender
}

\author{Goutam Das$^{1}$ and Daigo Shishika$^{2}$
\thanks{We gratefully acknowledge the support of ARL grant ARL DCIST CRA W911NF-17-2-0181.}
\thanks{$^{1}$Goutam Das is a PhD student of Electrical and Electronics Engineering,
        George Mason University, 4400 University Dr, Fairfax, VA 22030, USA
        {\tt\small gdas@gmu.edu}}%
\thanks{$^{2}$Daigo Shishika is an Assistant Professor of the Department of Mechanical Engineering, George Mason University,
        4400 University Dr, Fairfax, VA 22030, USA
        {\tt\small dshishik@gmu.edu}}%
}

\begin{document}

\maketitle
\thispagestyle{empty}
\pagestyle{empty}


\begin{abstract}
In this paper we consider a Target-guarding differential game where the Defender must protect a linearly moving line segment by intercepting the Attacker who tries to reach it.
In contrast to common Target-guarding problems, we assume that the Defender is attached to the Target and moves along with it.
This assumption affects the Defender's maximum speed depending on its heading direction.
A zero-sum differential game of degree for the Attacker-winning scenario is studied, where the payoff is defined to be the distance between the two agents at the time of reaching the Target.
We derive the equilibrium strategies and the Value function by leveraging the solution for the infinite-length Target scenario. 
The zero-level set of this Value function provides the barrier surface that divides the state space into Defender-winning and Attacker-winning regions.
We present simulation results at the end to demonstrate the theoretical results.
\end{abstract}







\section{INTRODUCTION}
Pursuit-evasion games (PEG) have been studied in different areas in robotics and controls community for various applications including missile guidance \cite{Garcia2017}, aircraft defense \cite{Fang2018, Vitaly2010}, robot navigation\cite{Guilamo2004}, self-driving vehicles \cite{Wang2021}, just to name a few.
This paper is interested in a particular class of PEG that involves a Target that must be guarded.
Such a scenario has high relevance to both civilian and military defense applications.


Target-Attacker-Defender (TAD) games
study situations where the Attacker seeks to reach the Target without being intercepted by the Defender.
The Target is modeled as a point/agent that is stationary \cite{Selvakumar2021}, or cooperates with the Defender by actively evading the Attacker or by rendezvousing with the Defender \cite{Liang2019,Liang2020,Liang2021}.
Defender can win either by intercepting the Attacker \cite{Garcia2017, Garcia2018, Garcia2019, Rubinsky2014}, or by rendezvousing with the Target \cite{OYLER2016}.

A related class of PEG is called the Target Guarding problem, which was first introduced by Isaacs \cite{Issacs1965}.
The main difference with the TAD game is that the Target is now a region instead of a point, which makes the rendezvous-type strategy to be invalid for the Defender.
There are many different variants of the Target Guarding problem including \emph{reach-avoid} game \cite{Zhou2012,Huang2011,Chen2014} and \emph{coastline guarding} or \emph{border-defense} problems\cite{Garcia_Coastline2019,VonMoll2020BD,Garca2020TheBS}.
These works extend the problem to multi-agent scenarios and consider different geometric settings, however,
it is generally assumed that the agents have simple motion and can freely move on a planar region.

We are interested in a Target Guarding scenario where the Defender is constrained to move along the perimeter of the Target.
Closely related works are previously studied as \emph{perimeter-defense} games \cite{VonMol2020, Shishika2020review,Shishika2020}.
Unlike the standard Target Guarding problems, it is assumed that the Defender cannot pass through the Target region. 
Therefore, the Defender must move around the perimeter to reach the Attacker, which affects the dynamics and thus the capturability.
Different variants have been studied with differential game technique \cite{VonMol2020} and with geometric approach \cite{Shishika2020review,Shishika2020}.
However, these works all considered stationary Target region.
In this paper, 
we consider a Defender that is constrained to move on the perimeter of a Target that moves in the space, which is relevant to convoy protection.
The Attacker moves freely and tries to reach the Target while avoiding the Defender. 
Note that, in the global frame, the Defender is dragged to the direction of Target's motion, but the Attacker is not affected by the motion of the Target. 
In this context, there is a connection to the work presented in \cite{Sun2017,Sun2015}, where PEG is played in a flow field.
However, the result does not extend naturally to our problem since we consider Target Guarding.
Moreover, in our problem, the flow field affects only one of the two agents.

The main contributions of the paper are: 
(i) the formulation of a new type of Target Guarding problem where the Defender must protect the perimeter of a moving Target; 
(ii) the characterization of the Barrier surface that separates the state space into Defender winning and Attacker winning regions; and
(iii) the equilibrium strategies and the Value function for the Attacker-winning scenario.
By considering the Target to be a line segment that translates in one direction, this paper serves as an initial step towards studying strategies to defend Targets with more complex shapes and motions.
\section{PROBLEM FORMULATION}\label{sec:prob_formulation}
\begin{figure}[tp]\label{fig_1}
    \centering
    \includegraphics[width =  \columnwidth]{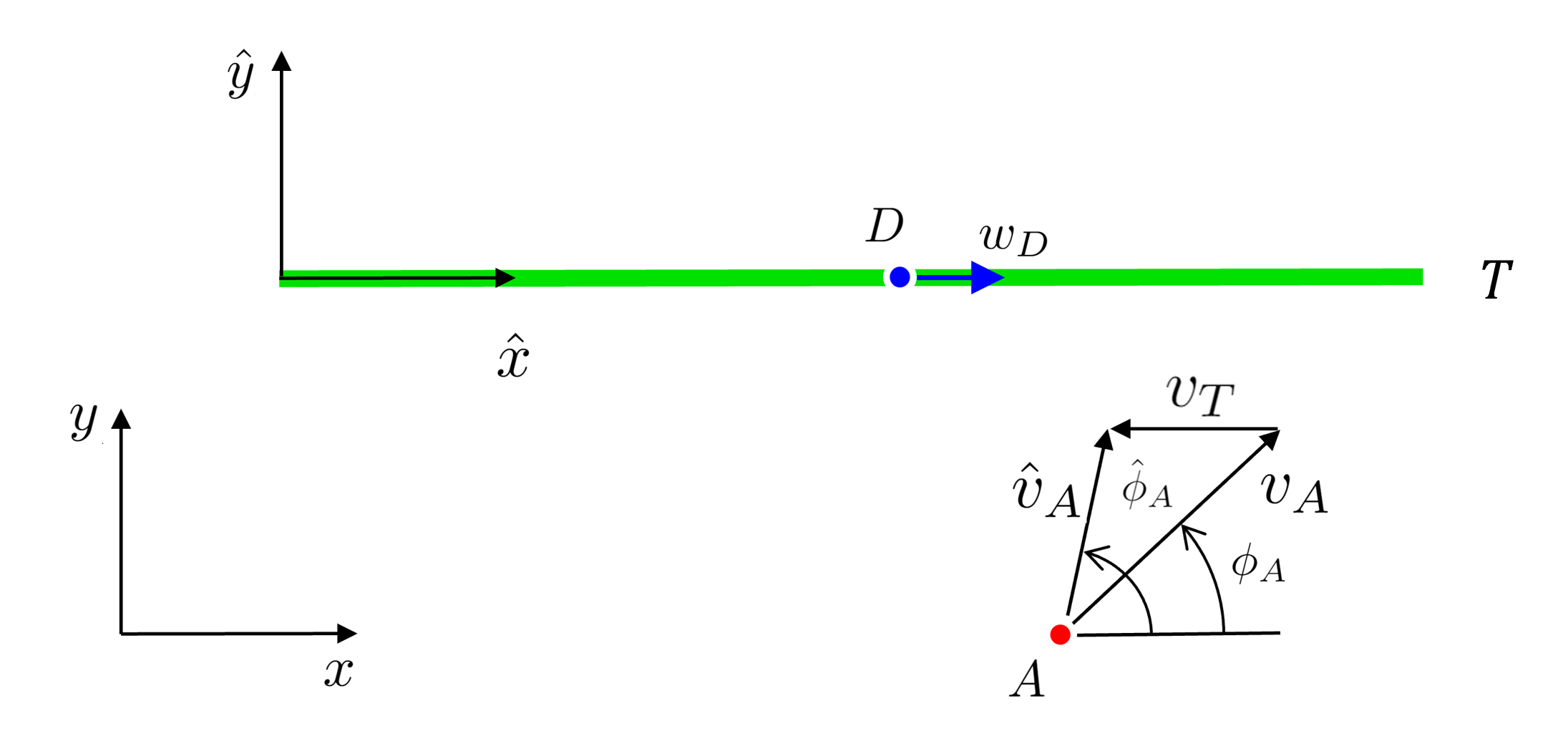}
    \caption{Illustration of guarding a translating linear Target with one Defender.}
    \label{fig:fig1}
\end{figure}

This section formulates the moving-Target Guarding problem with Defender attached on the Target. 
Figure~\ref{fig:fig1} shows the local frame that is attached to the Target with $(\hat{x}, \hat{y})$-coordinate system, as well as the inertial frame with $(x, y)$-coordinate system. 
The Target, 
$T$, is a line segment aligned with the $x$-axis with length $L$,  and it is moving at a constant speed, $v_T$, in the positive $x$ direction:
\begin{equation}\label{Target_dyn}
\begin{aligned}
    \begin{bmatrix}
    \dot{x}_T \\
    \dot{y}_T
    \end{bmatrix} &=     
    \begin{bmatrix}
     v_T \\
    0
    \end{bmatrix}.
\end{aligned}
\end{equation}
The Defender, $D$, is constrained to move on 
$T$, and it has first-order dynamics relative to 
$T$.
We use $[x_D, y_D]$ and $[\hat{x}_D, \hat{y}_D]$ to denote the position of the Defender in the inertial frame and moving frame, respectively.
Note that we have $\hat{x}_D= x_D - x_T$ and $\hat{y}_D=0$.
The Defender is able to move along $\tau$: i.e.,
\begin{equation}\label{Defender_dyn}
\begin{aligned}
    \begin{bmatrix}
    \dot{\hat{x}}_D \\
    \dot{\hat{y}}_D
    \end{bmatrix} &=     
    \begin{bmatrix}
    w_D \\
    0
    \end{bmatrix},
\end{aligned}
\end{equation}
where $w_D \in [-1, 1]$ is the Defender's control input.
Note that the Defender's speed is constrained in the moving frame: i.e., the maximum speed is specified relative to the Target. This assumption is motivated by defense mechanisms that are attached to the moving Target.
Therefore, the Defender's maximum speed in the inertial frame will either increase or decrease when the Defender is moving in either the same or opposite direction of the Target respectively. 

The Attacker, $A$, can move freely in $\RR^2$ with speed $v_A$:

\begin{equation}\label{Attacker_dyn}
\begin{aligned}
    \begin{bmatrix}
    \dot{x}_A \\
    \dot{y}_A
    \end{bmatrix} &=     
    \begin{bmatrix}
    v_A\cos{\phi_A} \\
    v_A\sin{\phi_A}
    \end{bmatrix},
\end{aligned}
\end{equation}
where $\phi_A \in [-\pi,\pi]$ is the Attacker's heading angle in the inertial frame, measured counterclockwise w.r.t. the positive $x$-axis.

Defining
    $\begin{bmatrix}
    \hat{x}_A \\
    \hat{y}_A
    \end{bmatrix}
    = 
    \begin{bmatrix}
    x_A \\
    y_A
    \end{bmatrix} - 
    \begin{bmatrix}
    x_T \\
    y_T
    \end{bmatrix}$, we have
\begin{equation}\label{Attacker_dyn_mf}
    \begin{aligned}
    \begin{bmatrix}
    \dot{\hat{x}}_A \\
    \dot{\hat{y}}_A
    \end{bmatrix} 
    &=
    \begin{bmatrix}
    \hat{v}_A\cos{\hat{\phi}_A}\\
    \hat{v}_A\sin{\hat{\phi}_A}
    \end{bmatrix}
    =     
    \begin{bmatrix}
    v_A\cos{\phi_A} - v_T\\
    v_A\sin{\phi_A}
    \end{bmatrix},
\end{aligned}
\end{equation}
where, $\hat{v}_A$ and $\hat{\phi}_A$ are the Attacker's speed and heading angle in moving frame respectively. 
Note that $\hat{v}_A$ is dependent on the heading angle $\phi_A$, and therefore, it is not a parameter.
Both the Attacker and the Defender are agile, i.e. they can move and change directions instantaneously.
The overall system dynamics in moving frame is as follows:
\begin{align}\label{System_dyn}
    \dot{\hat{\textbf{x}}} = f(\hat{\textbf{x}}) 
    =   \begin{bmatrix}
            \dot{\hat{x}}_D\\
            \dot{\hat{x}}_A\\
            \dot{\hat{y}}_A
        \end{bmatrix}
    =   \begin{bmatrix}
            w_D \\
            v_A\cos{\phi_A} - v_T\\
            v_A\sin{\phi_A}
        \end{bmatrix}.
\end{align}

In addition to $|w_D|\leq 1$, we make the following assumptions on the agents' speeds: 
\begin{itemize}\setlength{\itemindent}{1em}
    \item[\textbf{A1})] The Attacker is faster than the Target: $v_A>v_T$.
    \item[\textbf{A2})] The players' speeds are such that: $1-v_T>v_A$. 
\end{itemize}
The first assumption (A1) ensures that the Attacker can reach the Target starting from $\hat{x}_A(0)<0$.
Secondly, noting that $\dot{x}_D=w_D+v_T\in[-1+v_T,1+v_T]$, the second assumption (A2) ensures that the Defender can outrun the Attacker in the $x$ direction.
This implies that
the game is over (with Defender's win) if $\hat{x}_D = \hat{x}_A$ is achieved at some point in time, since the Defender has sufficient control authority to maintain $\hat{x}_A = \hat{x}_D$ regardless of the Attacker's control. 
Therefore, we consider $\hat{x}_A = \hat{x}_D$ to be part of the terminal condition: i.e., the Defender has successfully intercepted the Attacker and thwarted the attack. 

Finally, due to the symmetry, we make the following assumption on the initial condition:
\begin{itemize}\setlength{\itemindent}{1em}
    \item[\textbf{A3})] 
The initial Attacker position is such that $\hat{y}_A(0)<0$.
\end{itemize}

We define the \textit{Game of Kind} as the question of whether the Attacker can reach the Target with a non-zero distance from the Defender, or the Defender can prevent that by capturing (matching its $x$ coordinate with) the Attacker. In the following sections, the Barrier surface that separates these two cases is derived by solving a related \textit{Game of Degree}.


\subsection{Game of Degree}
To find the Barrier surface, we define a \emph{Game of Degree} which takes place when the Attacker reaches the Target before being intercepted by the Defender (i.e. $\hat{x}_A(t_f) \neq \hat{x}_D(t_f)$). In this case, the initial condition of the system lies in $\mathcal{R_A}$, which is the region of win for the Attacker. The game terminates when the Attacker reaches the Target, $T$.
The terminal condition is
\begin{align}\label{terminal_cond}
   \psi(\hat{\textbf{x}}_f,t_f) = \hat{y}_A(t_f) &= 0, \quad \forall  \hat{x}_A(t_f) \in [0, L],
\end{align}
where $t_f$ is the terminal time. 

We consider a zero-sum differential game with the following payoff that describes the \emph{miss distance}: 
\begin{equation}\label{payoff}
    \begin{aligned}
    J &= \Phi(\hat{\textbf{x}}_f,t_f) \\
      &= |{\hat{x}_A(t_f) - \hat{x}_D(t_f)}|.
\end{aligned}
\end{equation}
Here, the Defender is the minimizing player, and the Attacker is the maximizing player. The Attacker tries to reach the Target while maximizing the distance from the Defender at final time. On the other hand, the Defender wants to minimize $J$, i.e., it seeks to get as close as possible to the Attacker at final time. If an equilibrium exists, the Value function is defined as
\begin{align}
    V(\hat{\textbf{x}}) &= \min\limits_{w_D} \max\limits_{\phi_A} J = \max\limits_{\phi_A} \min\limits_{w_D} J.
\end{align}
where, $\hat{\textbf{x}} = [\hat{x}_D, \hat{x}_A, \hat{y}_A]$.

The equilibrium strategies $w_D^*$ and $\phi_A^*$ satisfy the following saddle-point condition:
\begin{align}
    J(w_D^*,\phi_A) \leq J(w_D^*,\phi_A^*) \leq J(w_D,\phi_A^*).
\end{align}

Here, we can consider the value function, $V$, to measure the level of performance for the Attacker and the Defender for a given initial condition. 
If $V>0$, it means that the Attacker can ensure a positive miss distance at the time of breaching regardless of the Defender strategy:  $\hat{\mathbf{x}}\in \mathcal{R}_A$.
In the critical case where $V=0$, the Defender has a strategy, $w_D^*$, to achieve zero distance (i.e., capture) at the time Attacker reaches the target.
Beyond this critical case, the Defender has a strategy to win the game, and the states are in the Defender winning region: $\hat{\mathbf{x}}\in \mathcal{R}_D$.
Note that the payoff, $J$, is invalid for the Defender-winning scenario, but it does not affect the critical case where $V=0$.

Based on the above properties, we will find the barrier surface, $\mathcal{B}$, as the zero level set of the Value function:
\begin{equation} \label{eq: Barrier}
    \mathcal{B} = \{\hat{\textbf{x}}\hspace*{1mm} | \hspace*{1mm} V(\hat{\textbf{x}})=0\}.
\end{equation}
We will derive $V$ and the corresponding equilibrium strategies in the following sections.


\section{INFINITE LENGTH TARGET}
As a building block towards the complete solution, we first assume the length of the Target to be infinite. The motion of the Target still affects the game through the speed that the Defender can achieve in the inertial frame (or equivalently, the relative speed that the Attacker can achieve in the moving frame).

For this alternate version of the problem, the Attacker can win the game by reaching the following \textit{terminal surface}, $\mathcal{S_T}$, which is the $\hat{x}$-axis of the moving frame:
\begin{align}\label{terminal_surface}
    \mathcal{S_T} = \{\hat{\textbf{x}} \; \mid \; \hat{y}_A(t_f) = 0\}.
\end{align}
Therefore, the terminal constraint is given by
\begin{align}\label{terminal_cond_inf}
   \psi(\hat{\textbf{x}}_f, t_f) = \hat{y}_A(t_f) &= 0. 
\end{align}
The assumptions (A1)-(A3) are retained.

\subsection{First Order Necessary Conditions for Optimality} 
The Hamiltonian for the differential game is defined as,
\begin{equation}\label{hamiltonian}
    \begin{aligned}
    \mathcal{H} &=  
    \sigma_{\hat{x}_D} w_D + 
    \sigma_{\hat{x}_A} v_A\cos{\phi_A} - \sigma_{\hat{x}_A} v_T + \sigma_{\hat{y}_A} v_A\sin{\phi_A} ,
\end{aligned}
\end{equation}
where $\sigma = [\sigma_{\hat{x}_D} \hspace{3mm} \sigma_{\hat{x}_A} \hspace{3mm} \sigma_{\hat{y}_A}]^\top$ is the adjoint vector.
The Hamiltonian is a separable function of the controls $w_D$ and $\phi_A$, and thus \textit{Isaacs’ condition} \cite{Issacs1965} , \cite{Basar2011} holds:
\begin{align}
    \min\limits_{w_D(t)} \max\limits_{\phi_A(t)} \mathcal{H} = \max\limits_{\phi_A(t)} \min\limits_{w_D(t)} \mathcal{H}.
\end{align}
The equilibrium adjoint dynamics are given by
    \begin{align}
    \dot{\sigma}_{\hat{x}_D} &= \frac{\partial \mathcal{H}}{\partial \hat{x}_D} = 0, \label{eq: sigma_xD_dot}\\
    \dot{\sigma}_{\hat{x}_A} &= \frac{\partial \mathcal{H}}{\partial \hat{x}_A} = 0, \label{eq: sigma_xA_dot}\\
    \dot{\sigma}_{\hat{y}_A} &= \frac{\partial \mathcal{H}}{\partial \hat{y}_A}
    = 0. \label{eq: sigma_yA_dot}
\end{align}

The terminal adjoint values are obtained from the transversality condition \cite{Bryson1975}:

\begin{equation} \label{eq: sigma_Transpose}
    \begin{aligned}
    \sigma^\top(t_f) &= 
    \begin{bmatrix}
    \sigma_{\hat{x}_D}(t_f) & \sigma_{\hat{x}_A}(t_f) & \sigma_{\hat{y}_A}(t_f)
    \end{bmatrix}
    \\
    &=\frac{\partial \Phi}{\partial \hat{\textbf{x}}_f} + \eta \frac{\partial \psi}{\partial \hat{\textbf{x}}_f}   \\
            &= \begin{bmatrix}
              \frac{\hat{x}_D- \hat{x}_A}{|\hat{x}_A - \hat{x}_D|} & \frac{\hat{x}_A - \hat{x}_D}{|\hat{x}_A - \hat{x}_D|} & 0
              \end{bmatrix} +
              \eta
              \begin{bmatrix}
              0 & 0 & 1
              \end{bmatrix}
    \\
    &=  \begin{bmatrix}
    -\lambda & \lambda & \eta
    \end{bmatrix},
\end{aligned}
\end{equation}

where
\begin{equation}
    \lambda := \text{sign}( \hat{x}_A- \hat{x}_D).
\end{equation}
Therefore, with \eqref{eq: sigma_xD_dot}-\eqref{eq: sigma_Transpose} we have the following:
\begin{equation}
    \begin{aligned}
    \sigma_{\hat{x}_D}(t) &= -\lambda,   &\forall \hspace{0.0mm} t \in [t_0, t_f]
    \\
    \sigma_{\hat{x}_A}(t) &= \lambda,    &\forall \hspace{0.0mm} t \in [t_0, t_f]
    \\
    \sigma_{\hat{y}_A}(t) &= \eta, &\forall \hspace{0.0mm} t \in [t_0, t_f].
\end{aligned}
\end{equation}
The terminal Hamiltonian satisfies,
\begin{equation}\label{eq: Hamiltonian_terminal}
     \begin{aligned}
     \mathcal{H}(t_f) = -\frac{\partial \Phi}{\partial t_f} -\eta \frac{\partial\psi}{\partial t_f} &= 0,
 \end{aligned}
\end{equation}
and $\frac{d\mathcal{H}}{dt} = 0$. Therefore, $\mathcal{H}(t) = 0$ for all $t \in [t_0, t_f]$.

The equilibrium control actions of the Attacker and Defender maximize and minimize (\ref{hamiltonian}) respectively: $\mathcal{H}^* = \text{max}_{\phi_A} \text{min}_{w_D} \mathcal{H}$. 
For the saddle-point solution of the problem, we have
\begin{equation}\label{wd*}
    \begin{aligned}
    w_D^* &= \argminD_{w_D} \mathcal{H}
    \\ &= \argminD_{w_D} (\sigma_{\hat{x}_D}w_D) = -\text{sign}({\sigma_{\hat{x}_D}}) = \lambda,
    \end{aligned}
\end{equation}
\begin{equation}\label{phi_A*}
    \begin{aligned}
    \phi_A^* &= \argmaxD_{\phi_A} \mathcal{H}\\
    &= \argmaxD_{\phi_A} (\sigma_{\hat{x}_A} v_A\cos{\phi_A} + \sigma_{\hat{y}_A} v_A\sin{\phi_A}).
    \end{aligned}
\end{equation}
Solving (\ref{phi_A*}), we obtain
\begin{equation}\label{cosphi*}
    \begin{aligned}
    \cos{\phi^*_A} 
    &= \frac{\sigma_{\hat{x}_A}}{\sqrt{\sigma_{\hat{x}_A}^2 + \sigma_{\hat{y}_A}^2}}  
    = \frac{\lambda}{\sqrt{\eta^2 + 1}},
    \end{aligned}
\end{equation}
\begin{equation}\label{sinphi*}
    \begin{aligned}
    \sin{\phi^*_A} &= \frac{\sigma_{\hat{y}_A}}{\sqrt{\sigma_{\hat{x}_A}^2 + \sigma_{\hat{y}_A}^2}} = \frac{\eta}{\sqrt{\eta^2 + 1}}.
    \end{aligned}
\end{equation}
Recalling assumption (A3), we consider the case where the Attacker approaches the Target from below.
Since there is no incentive for the Attacker to increase its distance from the Target, we know that $\phi_A 
\in[0,\pi]$.
However, if $\eta < 0$, this implies $\sin{\phi_A^*}<0$ due to (\ref{sinphi*}).
Therefore, it must be the case that $\eta > 0$.

Substituting the equilibrium controls, (\ref{wd*}), (\ref{cosphi*}) and (\ref{sinphi*}), into the Hamiltonian, (\ref{hamiltonian}), and evaluating at \textit{terminal time} with adjoints gives
\begin{equation}\label{eq: H*_tf}
\begin{aligned}
    \mathcal{H}^*(t_f) = 0 =
    & -\lambda^2 + \lambda v_A\bigg(\frac{\lambda}{\sqrt{\eta^2 + 1}}\bigg)  \\  &- \lambda v_T + \eta  v_A\bigg(\frac{\eta}{\sqrt{\eta^2 + 1}}\bigg).
\end{aligned}
\end{equation}
Since $\eta > 0$, solving \eqref{eq: H*_tf} gives
\begin{equation} \label{eq: eta}
    \begin{aligned}
    \eta  &= \frac{\sqrt{(1 + \lambda v_T)^2 - v_A^2}}{v_A}.
    \end{aligned}
\end{equation}
\subsection{Solution Characteristics}
The retrograde equilibrium kinematics can be obtained by substituting the equilibrium controls, (\ref{wd*}), (\ref{cosphi*}) and (\ref{sinphi*}), along with the adjoints into (\ref{Attacker_dyn_mf}) which yields
\begin{equation}\label{xA_dot}
\begin{aligned}
\mathring{\hat{x}}_A &= -v_T + \frac{\lambda v_A^2}{1+\lambda v_T},
\end{aligned}
\end{equation}

\begin{equation}\label{yA_dot}
\begin{aligned}
\mathring{\hat{y}}_A &= v_A \sqrt{1-\bigg(\frac{v_A}{1+\lambda v_T}\bigg)^2},
\end{aligned}
\end{equation}
with the boundary condition
\begin{align}
    \hat{y}_A(t_f)=0. 
    \label{eq:boundary_condition}
\end{align}
Now, let $[X,Y]$ denote the relative position of the Attacker with respect to the Defender: i.e.,
\begin{equation}\label{X,Y}
    [X,Y] = [x_A-x_D, y_A-y_D] = [\hat{x}_A-\hat{x}_D, \hat{y}_A].
\end{equation}
From (\ref{xA_dot}), (\ref{yA_dot}) and (\ref{X,Y}), we have
\begin{equation}\label{m1}
    \begin{aligned}
    \frac{dY}{dX} = m_1 := \frac{v_A\sqrt{\rho^2 -v_A^2}}{\lambda v_A^2-\rho(\lambda+v_T)}, 
    \end{aligned}
\end{equation}
where $\rho := 1+ \lambda v_T$.
This implies that the equilibrium trajectories of the system in the $XY$-plane is given by straight lines:
\begin{equation}\label{eq: eq_traject}
    Y = m_1 X + C.
\end{equation}
Note that, the value of $m_1$ depends on $\lambda = \text{sign}(X)$.
Furthermore, notice that the assumption (A2) ensures that $m_1$ is well-defined. 
Specifically, (A2) ensures that: (i) the term inside the square root, $\rho^2-v_A^2$, is always positive, and (ii) the denominator does not become zero.

\begin{figure}[h!]
    \centering
    \includegraphics[width = \columnwidth ]{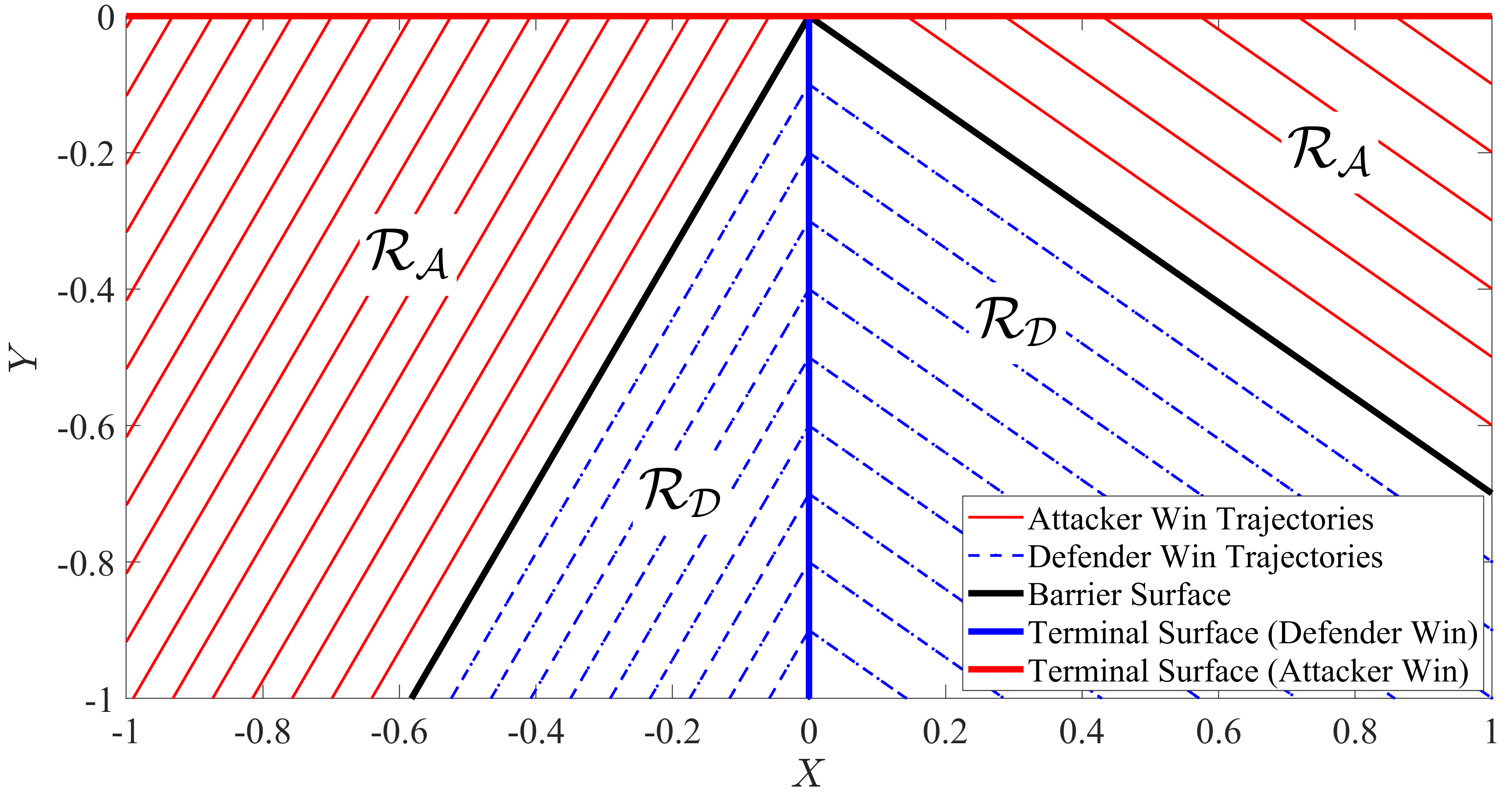}
    \caption{Full equilibrium flow-field with $v_A = 0.7$ and $v_T = 0.2.$}
    \label{fig:flow_field}
\end{figure}

The red solid lines in Fig.~\ref{fig:flow_field} presents the equilibrium trajectories for initial conditions satisfying (A3) and the boundary condition \eqref{eq:boundary_condition}.
The black solid line shows the critical case in which 
Attacker reaches the Target at the time of capture.
When the two agents use the same equilibrium strategies beyond this critical case, we obtain the set of trajectories shown in blue dashed lines.
Recalling that $X=0$ is part of the terminal surface corresponding to Defender's win, the trajectories terminate when they hit the $Y$-axis.


It can be seen that the terminal payoff in \eqref{payoff} is determined by the $X$ intercept of the state trajectory in the $XY$-plane, which we denote by $X_f$.
The Attacker tries to maximize $|X_f|$, while the Defender tries to minimize it (or possibly make the trajectory intersect the $Y$-axis).
The result of these two conflicting intentions is the equilibrium trajectories in Fig.~\ref{fig:flow_field}.

\def\mycmd{0}
\if\mycmd1
\color{red}
Now, suppose the Attacker is behind the Defender: i.e., $\lambda=-1$.
Then the Attacker's optimal trajectory in the inertial frame has the following slope:
\begin{equation}
    \frac{dy_A}{dx_A} = \frac{\sqrt{(1 - v_T)^2 - v_A^2}}{v_A}..
\end{equation}
If we consider the value of $v_A$ approach $(1-v_T)$ from below, the slope approaches 0 from the negative side.
This implies that, as $v_A$ increases, the Attacker will spend more effort in evading from the Defender (move in negative $x$ direction) than in approaching the Target (move in positive $y$ direction).
Importantly, the expression of the slope becomes invalid when $v_A>1-v_T$.
This degenerate scenario is explained in the following Lemma.

\begin{lm}\label{lem:attacker_speed}
Suppose the Target is infinitely long.
If $X<0$ and $v_A$ is greater than $1-v_T$: i.e., 
\begin{equation}
     v_A \in [1-v_T, 1],
\end{equation}
then the Attacker has a strategy to win the game against any admissible Defender strategy.
\end{lm}
\begin{proof}
In the inertial frame, the Defender's maximum speed moving in the negative $x$ direction is $-\dot{x}_D=1-v_T$, which is achieved when $w_D=-1$.

Now, if the Attacker has a maximum speed greater than $1-v_T$, and if it starts in a condition $X_0<0$ (i.e., $x_A(0)<x_D(0)$), then it can outrun the Defender moving towards the negative $x$ direction.
More formally, the Attacker has a strategy to achieve $\dot{X}=\dot{x}_A-\dot{x}_D<0$.
Therefore, for any given $Y_0<0$, the Attacker can increase $|X|$ arbitrarily large before it reaches the Target.
In the limiting case where $v_A=1-v_T$, the Attacker can move towards a point at $[-\infty,0]$ to achieve the payoff given by $X_0-\varepsilon$. 
\end{proof}

If the conditions in Lemma~\ref{lem:attacker_speed} are true (i.e., $X<0$ and $v_A>1-v_T$), then the game trivially ends with Attacker's win and the equilibrium strategies are not well defined.
The Attacker can choose to reach the Target after achieving arbitrarily large separation with the Defender.\footnote{Note that this will not be the case if the Target has a finite length, as is studied in Sec.~\ref{sec:finite_length_target}.}
Therefore, we exclude this case from the next theorem.
\color{black}
\fi

\begin{theorem}[Infinite-Length Target]\label{theorem: infinite}
Consider the Game of Degree with payoff given in \eqref{payoff}, and suppose the Target is infinitely long: i.e., spans the entire $\hat{x}$-axis.
Then the equilibrium state feedback control strategies are
\begin{equation}
\begin{aligned}\label{eq: eq_strategy_Attacker}
\big[\cos{\phi_A^*}, \sin{\phi_A^*}\big]
 &= \begin{cases}
 \dfrac{1}{\alpha}\Big[v_A, \sqrt{\alpha^2-v_A^2}\Big],
 \quad
 \hspace*{\fill} \text{if $
     X > 0 $; }
 \\[5pt]
 \dfrac{1}{\beta}\Big[-v_A, \sqrt{\beta^2-v_A^2}\Big],
 \quad 
 \hspace*{\fill} \text{if $
     X < 0 $;}
 \end{cases}
\end{aligned}
\end{equation} 
and
\begin{equation}
    \begin{aligned}
w_D^* &= \begin{cases}\label{eq: eq_strategy_Defender}
1,  \quad \hspace*{\fill} &\text{if $
     X>0$;}
\\
-1,  \quad \hspace*{\fill} &\text{if $
     X<0$;}
\end{cases}
\end{aligned}
\end{equation}
where $\alpha := 1+v_T$ and $\beta := 1-v_T$.
Moreover, the Value of the game is
\begin{equation}
    \begin{aligned}\label{Value_inf}
     V &= \begin{cases}
     X - \dfrac{Y}{m},
     
     \hspace*{\fill} & \text{if $X>0$; } 
     \\[10pt]
     -X + \dfrac{Y}{m},
     \hspace*{\fill} & \text{if $X<0$; }
     \end{cases}
\end{aligned}
\end{equation}
where 
\begin{equation}
    m(\phi_A^*,w_D^*) := \frac{v_A\sin{\phi_A^*}}{v_A \cos{\phi_A^*-w_D^*-v_T}}.
    \label{eq:optimal_slope_1}
\end{equation}

\end{theorem}
\begin{proof} The expressions for the Attacker's equilibrium strategy, $\cos{\phi_A^*}$ and $\sin{\phi_A^*}$, are obtained by substituting \eqref{eq: eta} into \eqref{cosphi*} and \eqref{sinphi*}. 
The Defender strategy, $w_D^*$, is given from \eqref{wd*}, which satisfies the first-order necessary condition for optimality. 

Given the strategies, the relative position takes a straight line path in the  $XY$-plane, $Y=mX+C$, with the slope given in \eqref{eq:optimal_slope_1}.\footnote{This is a generic expression of the slope for any given $\phi_A$ and $w_D$. One can verify that $m$ in \eqref{eq:optimal_slope_1} matches $m_1$ in \eqref{m1}, when the agents use the equilibrium strategies.}
As discussed with Fig.~\ref{fig:flow_field}, the Value is given by the $X$ intercept of the equilibrium trajectory.
More specifically, the \emph{miss distance} is $X_f>0$ if the game starts in the positive $X$ region, whereas it is $-X_f>0$ if the game starts in the negative region (and thus the $X$ intercept is negative).

For a given initial condition $[X_0, Y_0]$, we have
\begin{equation}
\begin{aligned}
  C   &= Y_0 - m X_0.
\end{aligned}
\end{equation}
Substituting $C$ back into the equation and solving for the $X$ intercept gives:
\begin{equation}\label{eq:x_f}
    X_f = X_0 - \frac{Y_0}{m}.
\end{equation}
This completes the proof that \eqref{Value_inf} provides the Value of the game.
\end{proof}

\section{FINITE LENGTH TARGET }
\label{sec:finite_length_target}

Now, consider the original problem where the length of the Target is $L$. The Defender, $D$, can travel only upto the end point of the Target which we denote by $E(\hat{x}_E,0)$, where $\hat{x}_E$ is defined as follows:
\begin{align}\label{eq: endpoint}
    \hat{x}_E &= (1+\lambda)\frac{L}{2}.
\end{align}

The Defender will follow the same strategy irrespective of the Target length since it only depends on the sign of $X$.
However, the strategy \eqref{eq: eq_strategy_Attacker} for the Attacker may not be valid, since it may not intersect the Target with finite length.
If the optimal strategy given by Theorem~\ref{theorem: infinite} does not intersect the Target, the Attacker must choose an alternate heading. 

Now, let $B(\hat{x}_B,0)$ denote the point on the $\hat{x}$-axis that the Attacker reaches following \eqref{eq: eq_strategy_Attacker}:
\begin{equation}\label{hat(x)_B}
    \hat{x}_B := \hat{x}_A - \frac{\hat{y}_A}{m_B},
\end{equation}
where
\begin{equation}
    m_B := \frac{\sin\phi_A^*}{\cos\phi_A^*}=\frac{\sqrt{\rho^2-v_A^2}}{\lambda v_A}.
\end{equation}
Note that the optimal headings are given in \eqref{eq: eq_strategy_Attacker}.
The strategy in \eqref{eq: eq_strategy_Attacker} is still valid for finite-length Target case if and only if the following condition holds:
\begin{equation}\label{eq:xB_condition}
    \hat{x}_B \in [0,L].
\end{equation}

If \eqref{eq:xB_condition} does not hold, then the Attacker must sacrifice the separation with the Defender and pick an aim point that actually intercepts the Target.\footnote{Note that there is no incentive for the Attacker to go around the end point and approach the Target from the positive side: i.e., enter $\hat{y}_A>0$ region.}
The aim point that achieves the least deviation from \eqref{eq: eq_strategy_Attacker} is the end point $E$.

\begin{figure}[ht]
    \centering
    \includegraphics[width = 7cm]{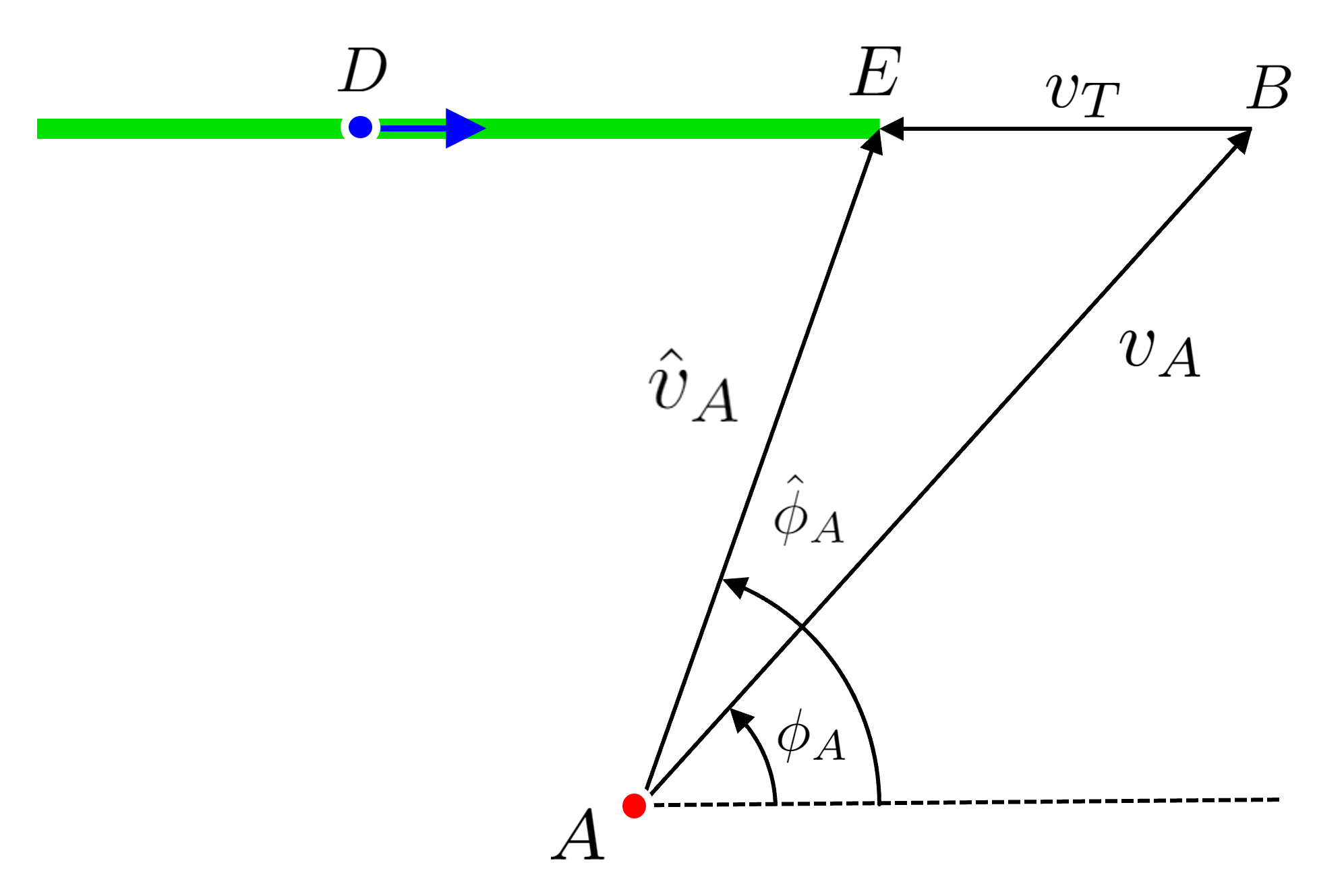}
    \caption{Attacker optimal heading for the end point of the Target in inertial and moving frame.}
    \label{Fig: A_Traject}
\end{figure}
When the Attacker aims for the end point, $E$, the desired heading angle in the moving frame can be obtained as follows
\begin{align}
    \cos{\hat{\phi}_A^\finite} &= \frac{\hat{x}_E-\hat{x}_A}
    {d_{EA}}, \label{eq:cos_fin}
    \\
    \sin{\hat{\phi}_A^\finite} &=
    \frac{-\hat{y}_A}
    {d_{EA}}, \label{eq:sin_fin}
\end{align}
where $d_{EA} := \sqrt{(\hat{x}_E-\hat{x}_A)^2+\hat{y}_A^2}$ is the distance from the Attacker's position, $A$, to Target's endpoint, $E$.
Note that we use the superscript $^\finite$ to denote the optimal strategies for the finite-length case.


Using the law of cosines for the triangle $\triangle ABE$ in Fig.~\ref{Fig: A_Traject}, we obtain
\begin{equation}\label{eq: law of cosine} 
 v_A^2 = v_T^2+\hat{v}_A^2-2v_T\hat{v}_A\cos({\pi -\hat{\phi}^\finite_A}).
\end{equation}
Solving for $\hat{v}_A$ yields
\begin{equation} 
    \begin{aligned}
    \hat{v}_A^\finite &= -v_T \cos \hat{\phi}_A^\finite + \sqrt{v_A^2 - v_T^2 \sin^2\hat{\phi}_A^\finite} 
    \\
    &= \frac{1}{d_{EA}}\bigg(-v_T(\hat{x}_E-\hat{x}_A)+\sqrt{(v_A d_{EA})^2 + (v_T \hat{y}_A)^2}\bigg). \label{eq:v_Ahat_fin}
\end{aligned}
\end{equation}

Now we are ready to state the main theorem.



\begin{figure}[H]
    \centering
    \includegraphics[width = \columnwidth]{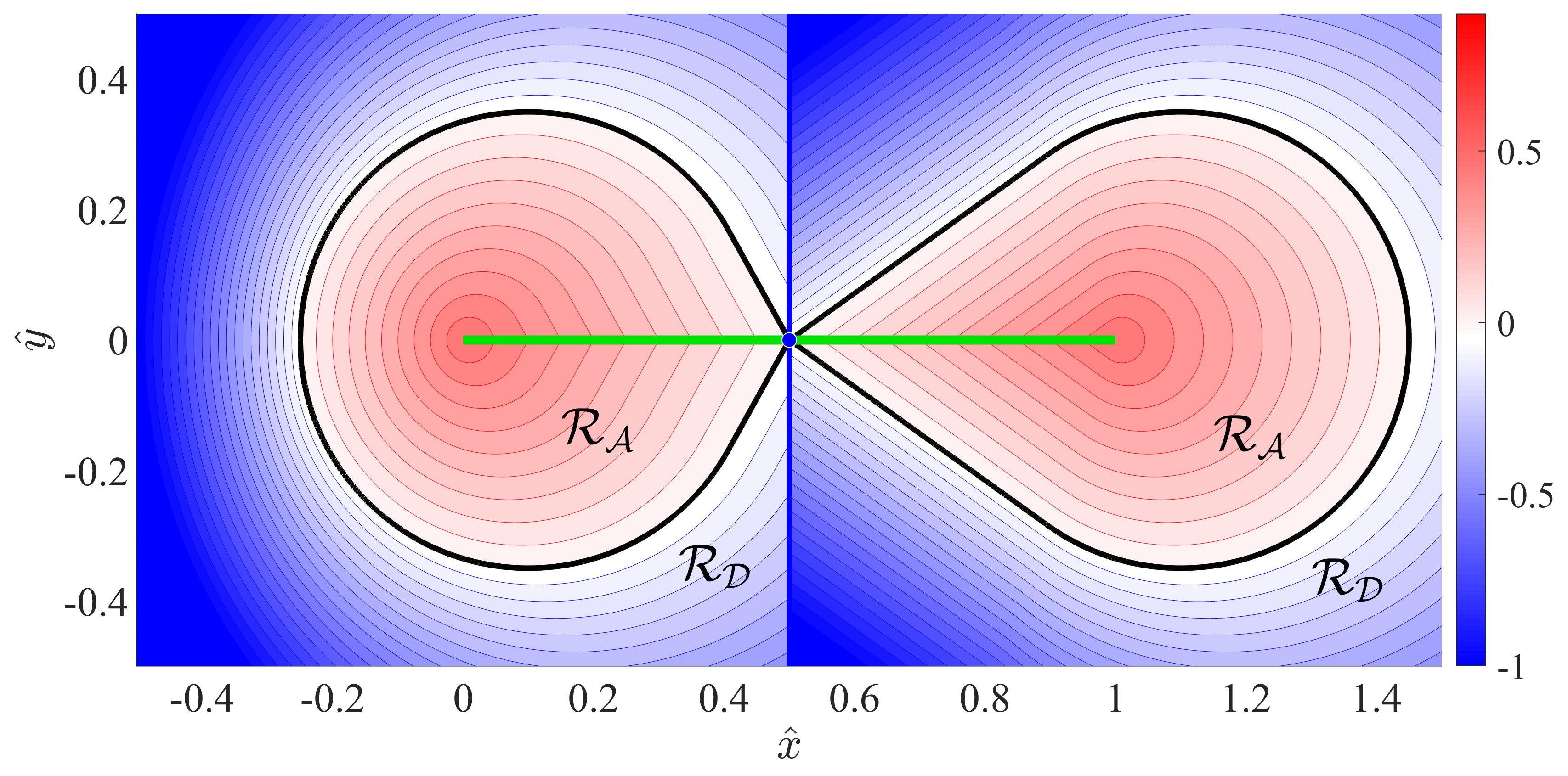}
    \caption{Level set of the Value functions when $\hat{x}_D = [0.5,0]$. The zero-level set provides the Barrier that separates the Attacker and Defender winning region. 
    The Defender's $\hat{x}$ position separates the space into two regimes: $X>0$ and $X<0$.
    }
    \label{fig:level_set}
\end{figure}

\begin{theorem}[Finite-Length Target] \label{theorem: finite}
The equilibrium state feedback control strategy for the Defender is given by
\begin{equation}
    w_D^\finite =
    \begin{cases}
 1,  \quad \hspace*{\fill} &\text{if $X>0$; }\\
-1,  \quad \hspace*{\fill} &\text{if $X<0$}.
    \end{cases}
\end{equation}
The equilibrium state feedback strategy for the Attacker is given by \eqref{eq: eq_strategy_Attacker} if condition \eqref{eq:xB_condition} holds; 
otherwise, it is given by
\begin{equation}\label{eq:optimal_A_finite}
    \big[\cos{\phi_A^\finite}, \sin{\phi_A^\finite}\big]
 = \dfrac{1}{v_A}\Big[\hat{v}_A^\finite \cos{\hat{\phi}_A^\finite}+v_T, \hat{v}_A^\finite\sin{\hat{\phi}_A^\finite}\Big].
\end{equation}
where $\cos{\hat{\phi}_A^\finite}$, $\sin{\hat{\phi}_A^\finite}$, and  $\hat{v}_A^\finite$ are given by \eqref{eq:cos_fin}, \eqref{eq:sin_fin} and \eqref{eq:v_Ahat_fin} respectively.
The Value of the Game is
given by the expression in \eqref{Value_inf} and \eqref{eq:optimal_slope_1}. 
Note that based on condition \eqref{eq:xB_condition}, either \eqref{eq: eq_strategy_Attacker} or \eqref{eq:optimal_A_finite} is used in \eqref{eq:optimal_slope_1} to compute $m$.
     
\end{theorem}
\begin{proof}
The Defender's strategy is the same for both the finite and infinite length target, as it moves towards the Attacker as fast it can. 

For the Attacker, if the heading angle given by Theorem~\ref{theorem: infinite} does intersect the Target, then it is the best choice for the Attacker as they are derived from the necessary conditions for optimalilty. If not, the Attacker has to choose the heading angle so that it can reach the Target by deviating the least amount possible from the equilibrium strategies given by \eqref{eq: eq_strategy_Attacker}. 
Considering these two conditions, the endpoint of the Target given by \eqref{eq: endpoint} is the optimal choice for the Attacker.
The heading in \eqref{eq:optimal_A_finite} is obtained by substituting \eqref{eq:cos_fin}, \eqref{eq:sin_fin} and \eqref{eq:v_Ahat_fin} into \eqref{Attacker_dyn_mf}.

From \eqref{eq:x_f}, we have seen that for any given initial conditions and $m$, 
the expression in \eqref{Value_inf} gives the $X$ intercept (payoff).
Furthermore, the expression in \eqref{eq:optimal_slope_1} provides the slope for any given strategy.
Therefore, the generic expression \eqref{Value_inf} and \eqref{eq:optimal_slope_1} can be used for the finite-length strategy \eqref{eq:optimal_A_finite} as well.
\end{proof}
Now, for a given set of initial states, we can calculate the Value of the \emph{Game of Degree} and plot the level set.
Figure~\ref{fig:level_set} shows the winning regions and associated Value of the game given that $\hat{x}_D = [0.5,0], v_A = 0.7 \text{ and } v_T = 0.25$. The black line indicates the Barrier surface  for the given Defender position. 
The payoff in \eqref{payoff} is defined by the Attacker's distance from the Defender once it successfully reaches the Target. 
The red shaded zone shows the positive \textit{miss distance} for the Attacker and indicates the Attacker's winning region. 
The Barrier surface is shown by the black line where the Value is zero, i.e., for these set of initial conditions the Attacker will meet the Defender on Target with zero \textit{miss distance}.
For any states given in the blue region, 
the Defender achieves $X=0$ (or $\hat{x}_A=\hat{x}_D$) before the Attacker reaches the Target.
Although we have not defined the payoff for this Defender-wining scenario, the level set of $V$ illustrates how close the states are from the Barrier: i.e., the level of threat.

By deriving the Barrier surface with a zero level set of $V$, we solved the \textit{Game of Kind} while establishing the winning regions.
Additionally, the blue vertical line is the Terminal surface for the Defender winning case. 
This vertical line separates the equilibrium strategies into two cases: $X>0$ (or $\lambda=1$) and $X<0$ (or $\lambda=-1$).
\begin{figure}[H]
    \centering
    \includegraphics[width = \columnwidth]{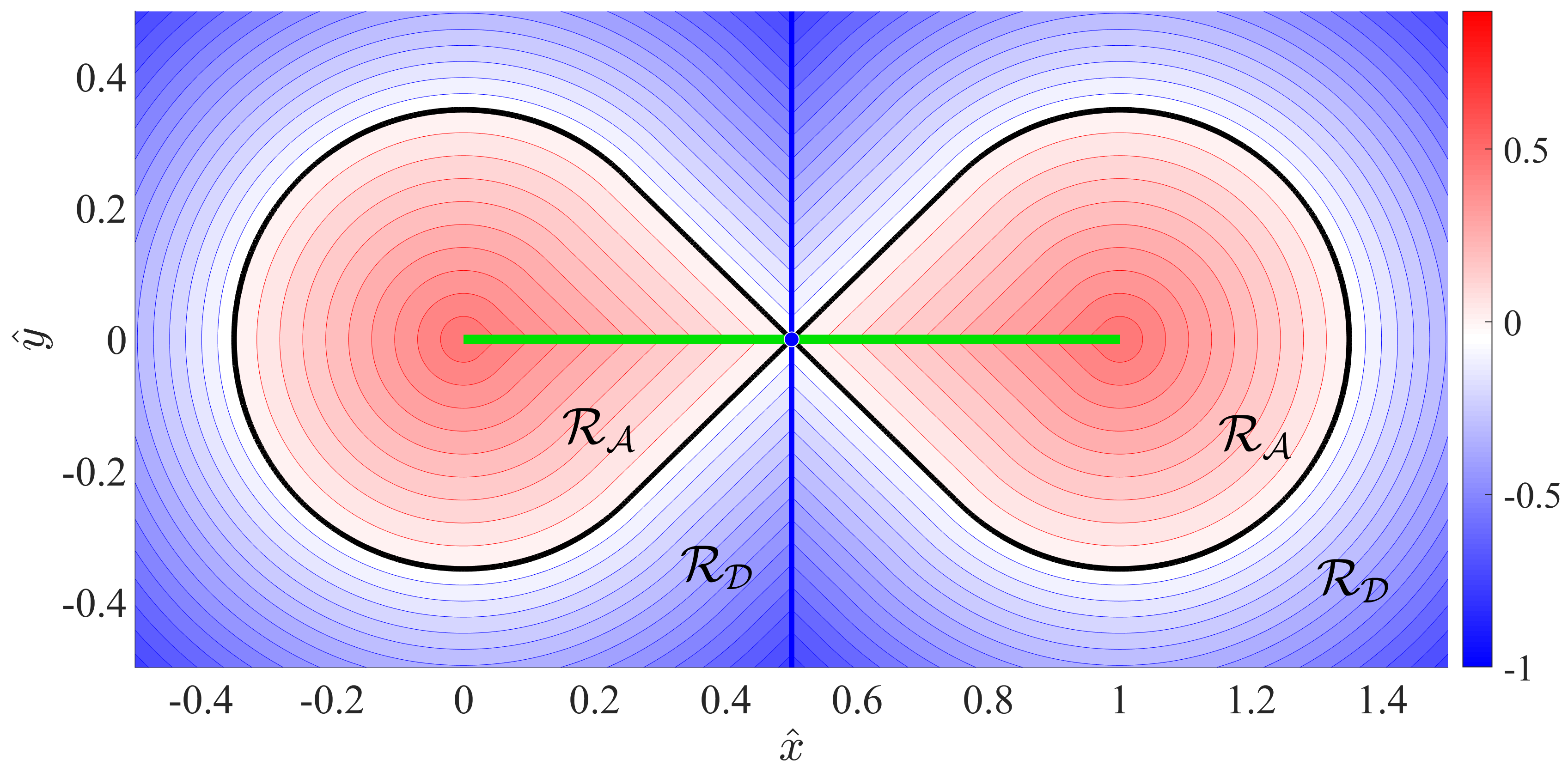}
    \caption{Level set of the Value functions the Attacker and Defender winning region for static Target with $v_A = 0.7$.}
    \label{fig:level_set_static}
\end{figure}
Finally, we present a degenerate case where the Target is static.
Figure~\ref{fig:level_set_static} shows the level sets for
the same parameters as in Figure~\ref{fig:level_set}, except now $v_T=0$.
In this case, the winning regions for the Attacker and the Defender are symmetric in both $\hat{x}$ and $\hat{y}$-axis given that the Defender's initial position is exactly at the middle of the Target.
Compare this to the case when the Target is moving along the $x$-axis shown in Figure~\ref{fig:level_set}.
We can see how the target motion affects the size and shape of the winning regions.

\section{SIMULATIONS}
We show both Attacker-winning and Defender-winning scenarios for the following parameters:
$v_A=0.60$, $v_T=0.35$  and $L=1$.\footnote{
The animated version of the simulations can be found online at \texttt{https://youtu.be/GW5CpHBT9oQ}}

\subsection{Attacker Winning Scenario}
\begin{figure}[!ht]
    \centering
    \includegraphics[width = \columnwidth]{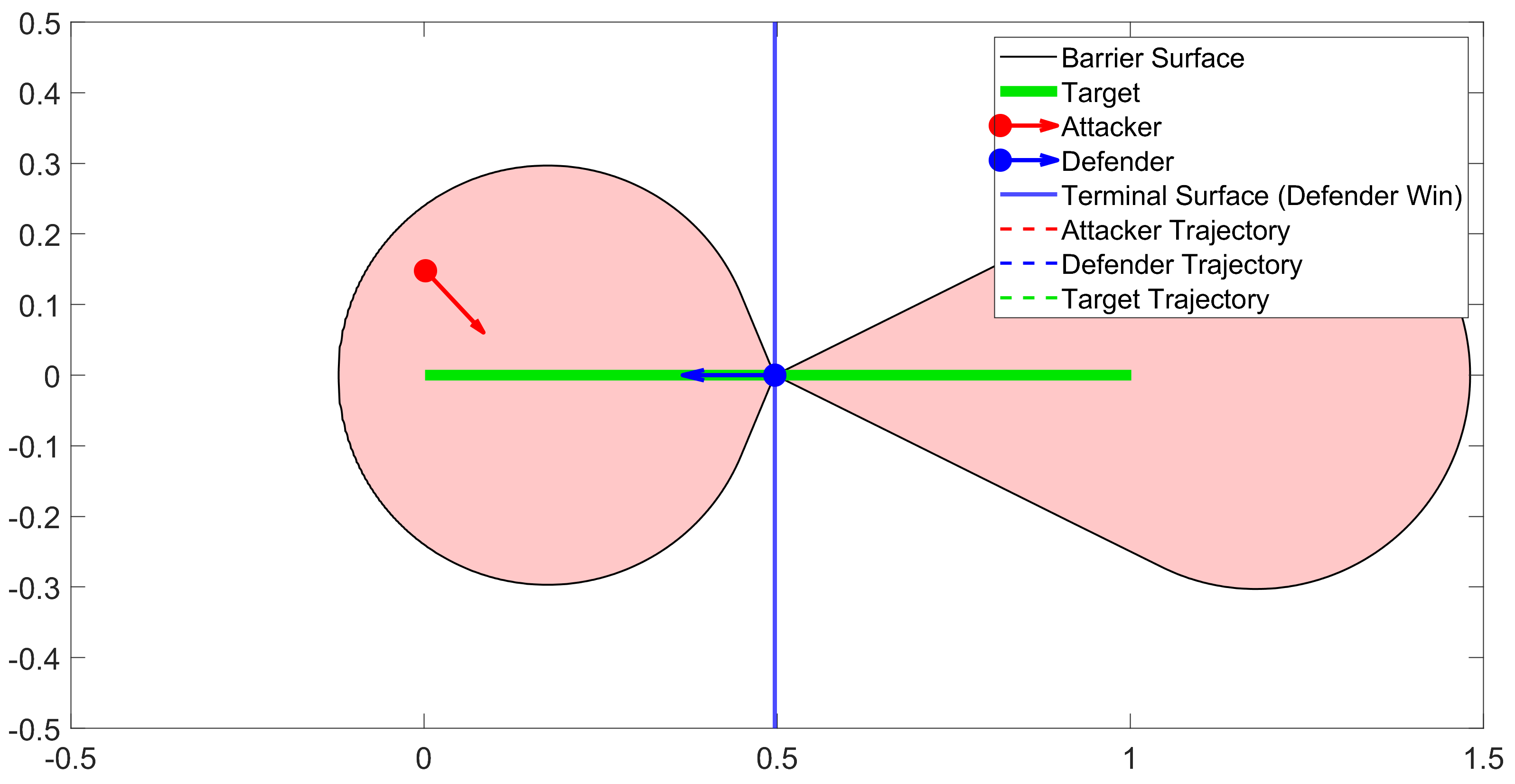}
    \caption{Initial position of the Defender and Attacker: $[\hat{x}_D,\hat{y}_D] = [0.5, 0]$ and $[\hat{x}_A,\hat{y}_A] = [0, 0.15]$ respectively. The arrows indicate the direction of movements for the Attacker and Defender.}
    \label{fig:Sim1_init}
\end{figure}

\begin{figure}[!ht]
    \centering
    \includegraphics[width = \columnwidth]{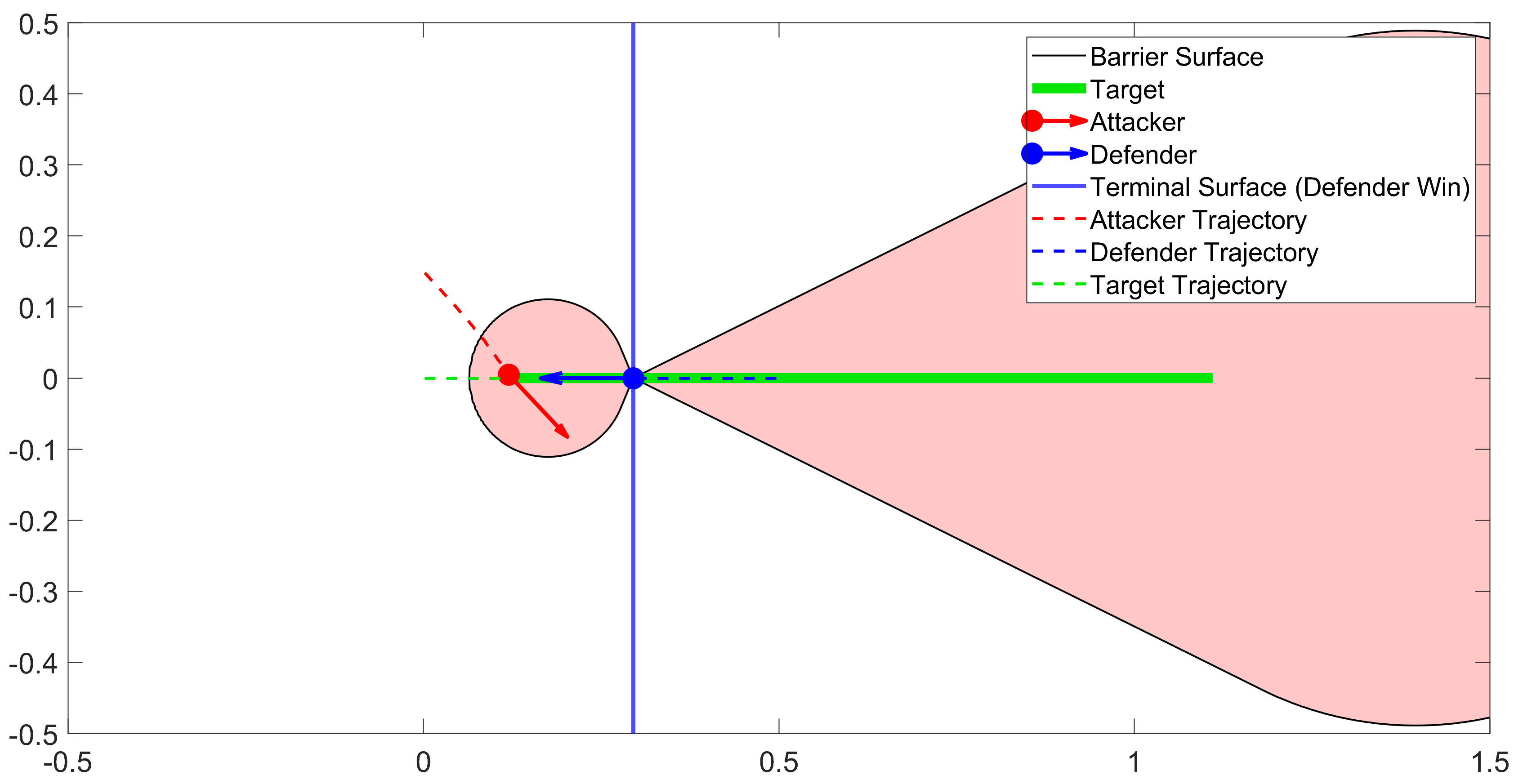}
    \caption{The Attacker reaches the Target and wins the Game. The dashed lines indicate the trajectories of the agents (Attacker, Defender and Target). Note that the Target also leaves a trail as it moves along the $x$-axis.}
    \label{fig:Sim1}
\end{figure}

The initial states are $[\hat{x}_D, \hat{y}_D] = [0.5, 0]$ and $[\hat{x}_A, \hat{y}_A] = [0, 0.15]$, which gives $V= 0.1922$. Figure~\ref{fig:Sim1_init} and Figure~\ref{fig:Sim1} show the initial and final state of the game respectively.
The shaded area indicates the Attacker winning region. 
In this case, the Attacker starts within that region and eventually win the game by reaching the Target without being intercepted by the Defender.

\subsection{Defender Winning Scenario}
\begin{figure}[!ht]
    \centering
    \includegraphics[width = \columnwidth]{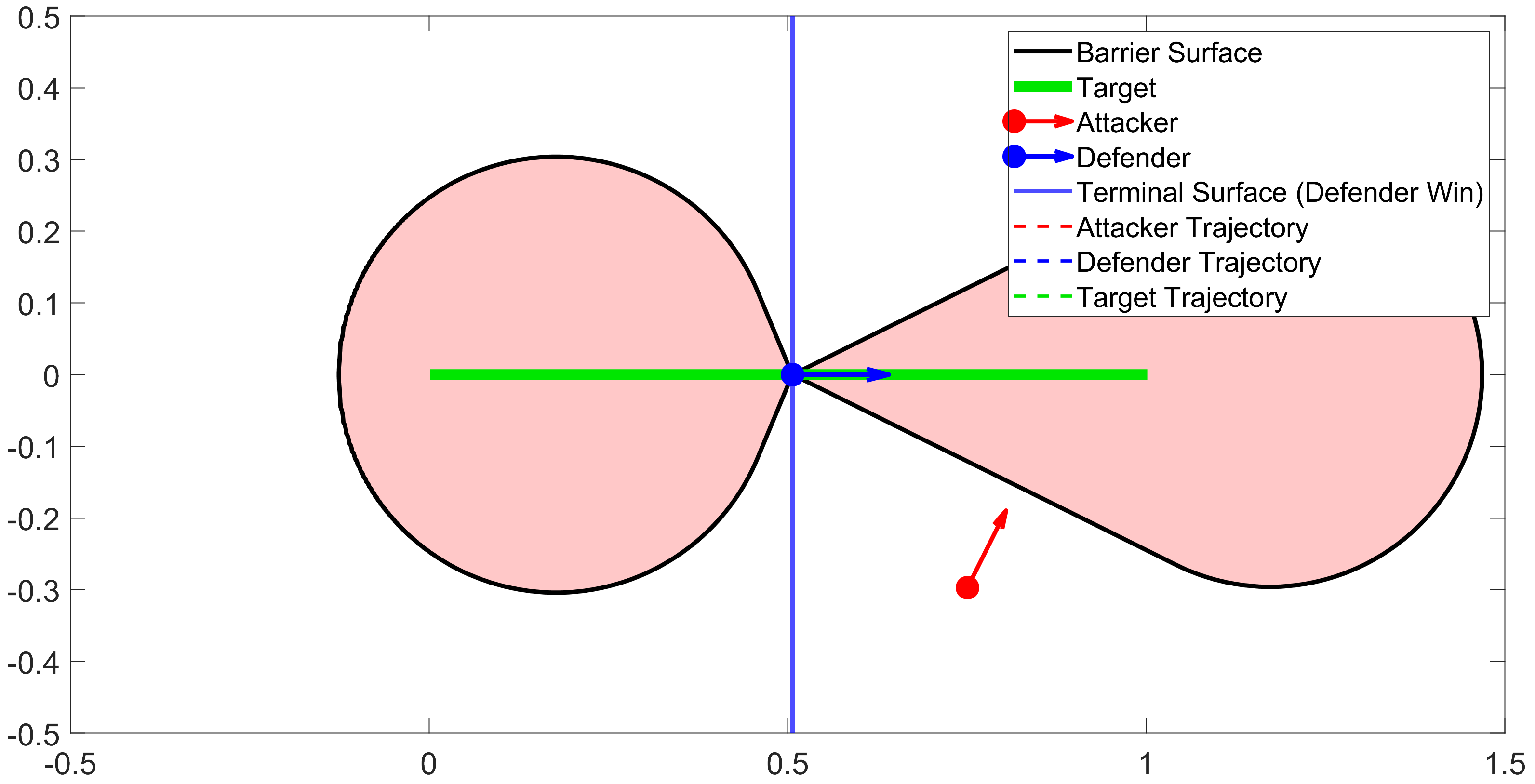}
    \caption{The Attacker starts outside of the winning region and the red arrow indicates the direction of the Attacker's motion heading towards the Target. The Defender also moves to the same direction of the Attacker in the $\hat{x}$-axis.}
    \label{fig:Sim2_init}
\end{figure}

\begin{figure}[!ht]
    \centering
    \includegraphics[width = \columnwidth]{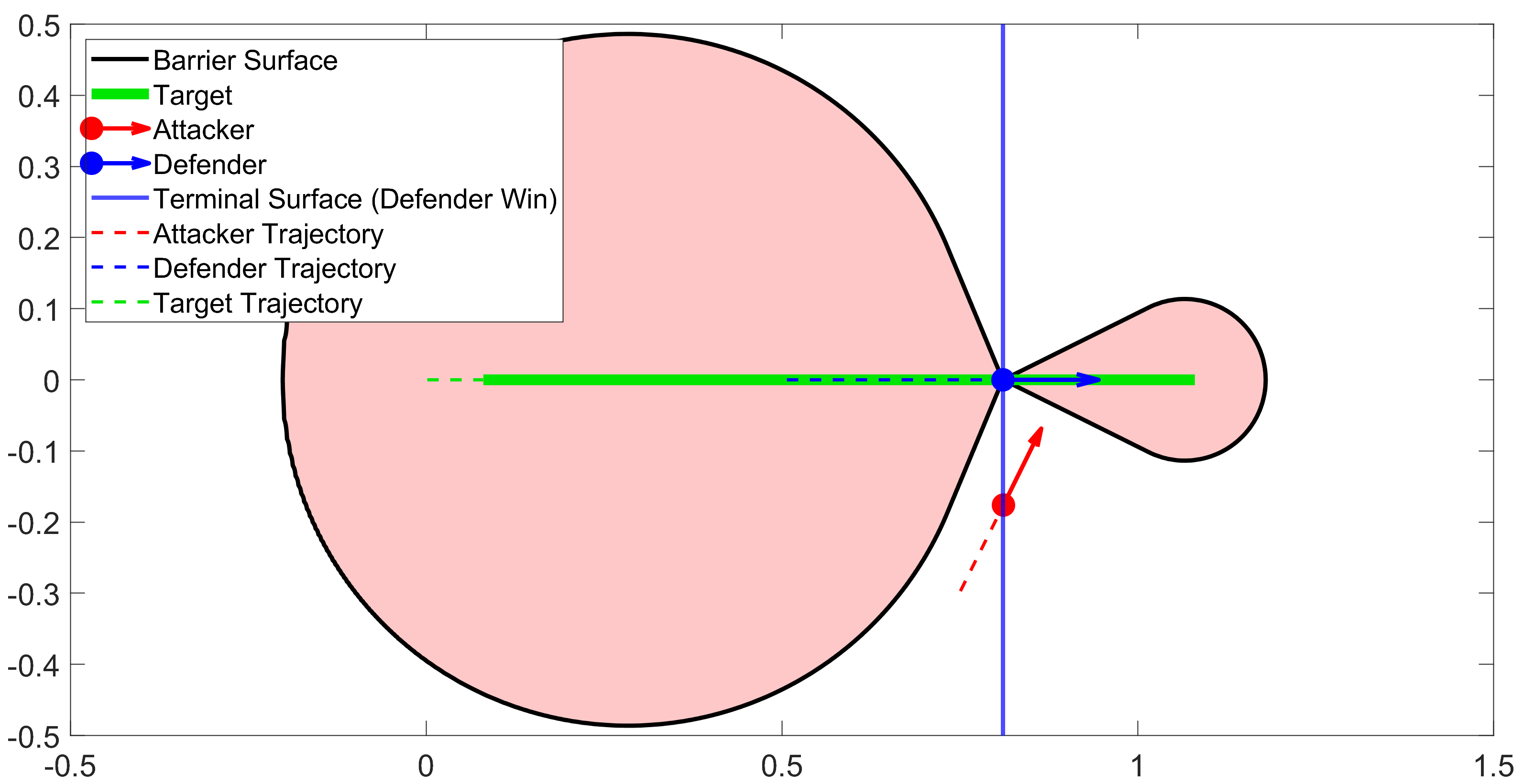}
    \caption{The Defender has successfully intercepted the Attacker. The blue vertical line shows that the Defender has aligned with the Attacker i.e., reaching the terminal condition for Defender winning Case.}
    \label{fig:Sim2}
\end{figure}

Now consider $[\hat{x}_D,\hat{y}_D] = [0.5, 0]$ and  $[\hat{x}_A,\hat{y}_A] = [0.75, -0.2]$, which gives $V = -0.1531$. The Attacker is outside of the Attacker winning region given by the shaded area. Figure~\ref{fig:Sim2_init} and \ref{fig:Sim2} show the initial and terminal states of the game. At the final time, the Defender intercepts the Attacker before reaching the Target and wins the game.


\section{CONCLUSIONS}
In this paper, the problem of guarding a moving Target is considered for one Defender and one Attacker scenario. 
For the Attacker-winning case, we formulated a \textit{Game of Degree} as a zero-sum differential game with a payoff defined as the distance between the players at the time of breaching. 
As a building block, we first solve the case when the Target is infinitely long.
The result is used to address the original problem where Target has a finite length.
We identify the equilibrium strategies and the Value function, whose zero-level set provides the Barrier surface for the \textit{Game of Kind}.
Future work on this problem may include different shapes and other possible motions of the Target.
Cooperative and active Target-defense for multiple Attackers and multiple Defenders can be another challenging problem to investigate.

\addtolength{\textheight}{-12cm}  

\section*{ACKNOWLEDGMENT}
Thanks to Dr. Cameron Nowzari for his recommendations and encouragement to initiate the research.

\bibliographystyle{IEEEtran}
\bibliography{citation}
\end{document}